\documentclass{PoS}

\newcommand {\apgt} {\ {\raise-.5ex\hbox{$\buildrel>\over\sim$}}\ }
\newcommand {\aplt} {\ {\raise-.5ex\hbox{$\buildrel<\over\sim$}}\ }

\title{Near-pristine gas at high redshifts: a window on early nucleosynthesis}

\ShortTitle{Near-pristine gas at high redshifts: a window on early nucleosynthesis}

\author{\speaker{Max Pettini} \\
        Institute of Astronomy, University of Cambridge\\        
        E-mail: \email{pettini@ast.cam.ac.uk}}

\author{Ryan Cooke\\
           Institute of Astronomy, University of Cambridge, and \\
           Department of Astronomy \& Astrophysics, University of California, Santa Cruz\\
           E-mail: \email{rcooke@ucolick.org}}

\abstract{It has now become recognised that
damped Lyman alpha systems---gas clouds of neutral hydrogen
observed in the high redshift Universe---play an important
role in helping us unravel the origin of chemical elements.
Here we describe the main results of a recently completed
survey of the most metal-poor DLAs, aimed at complementing and
extending studies of the oldest stars in the Galaxy.
The survey has clarified a number of lingering issues concerning
the abundances of C, N, O in the low metallicity regime,
has revealed the existence of DLA analogues to
Carbon-enhanced metal-poor stars, and is providing some
of the most precise measures of the primordial abundance
of Deuterium.
}

\FullConference{XII International Symposium on Nuclei in the Cosmos\\
                 August 5-12, 2012\\
                 Cairns, Australia}

\begin{document}

\section{Introduction}

One of the recurrent themes at this meeting has been
the search for, and the study of, the most metal-poor 
stars in our Galaxy and its neighbours. The ultimate 
goal of such searches is to dig into the chemical
composition of these fossil remnants for clues to the
nature of the  `First Stars' which reionised the Universe 
and seeded it with the first generation of elements
heavier than Li.

\begin{figure}[h!]
\medskip
\centerline{~~~\includegraphics[width=0.75\textwidth]{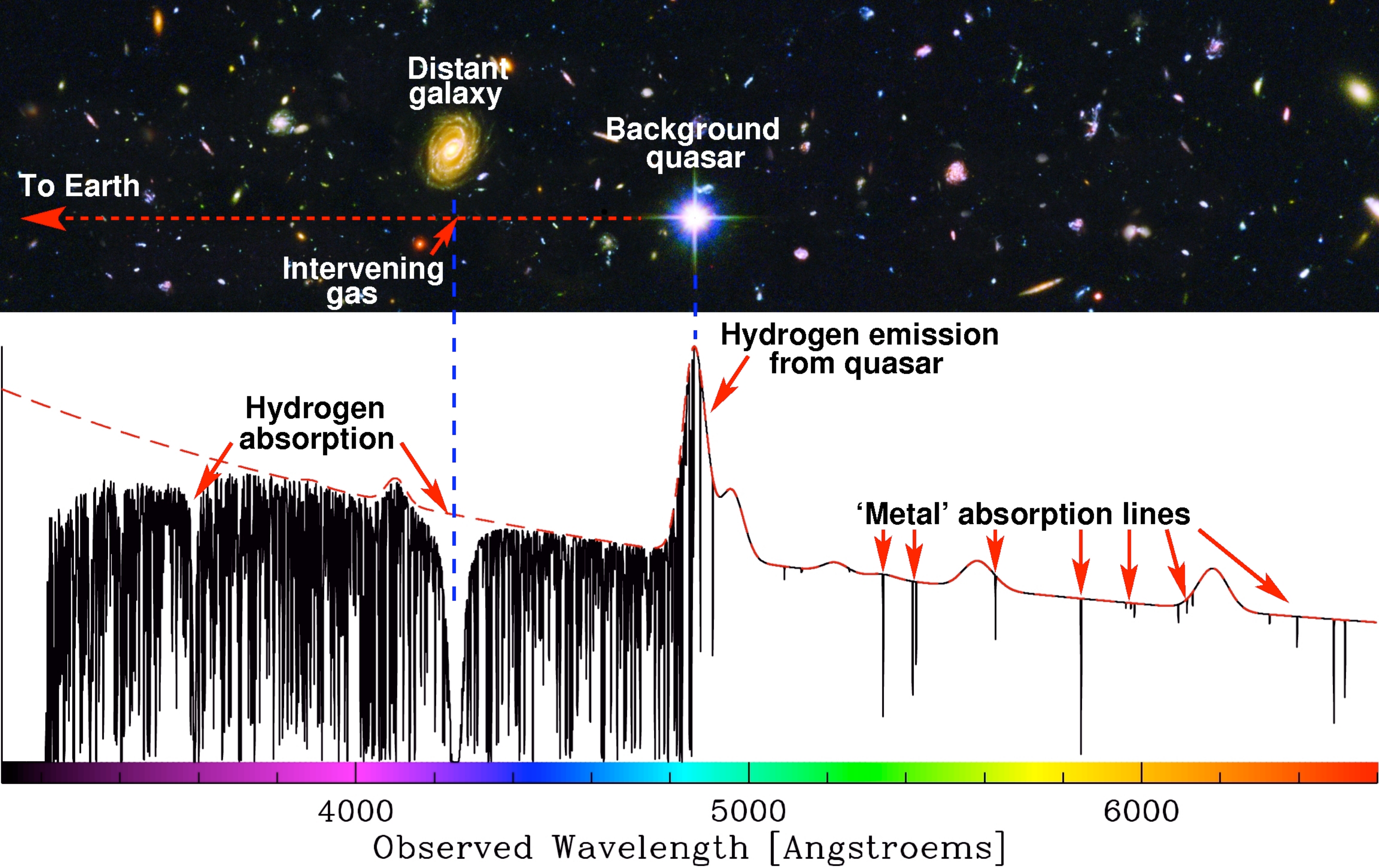}}
\caption{An illustration of the quasar absorption line 
technique used to probe the distant Universe 
(courtesy of J. Webb and M. Murphy).}
\label{fig:QSO_AbsLines}
\end{figure}

In parallel with these efforts, we have been
pursuing a complementary approach which aims
to identify near-pristine pockets of gas at high
redshift which may be the counterparts of the 
gas clouds from which the oldest Galactic stars 
condensed. The technique we have employed
in this endeavour,
called quasar (or QSO for short) absorption line spectroscopy,
 is illustrated in Figure~\ref{fig:QSO_AbsLines}.

\begin{figure}[]
\centerline{~~~\includegraphics[width=0.95\textwidth]{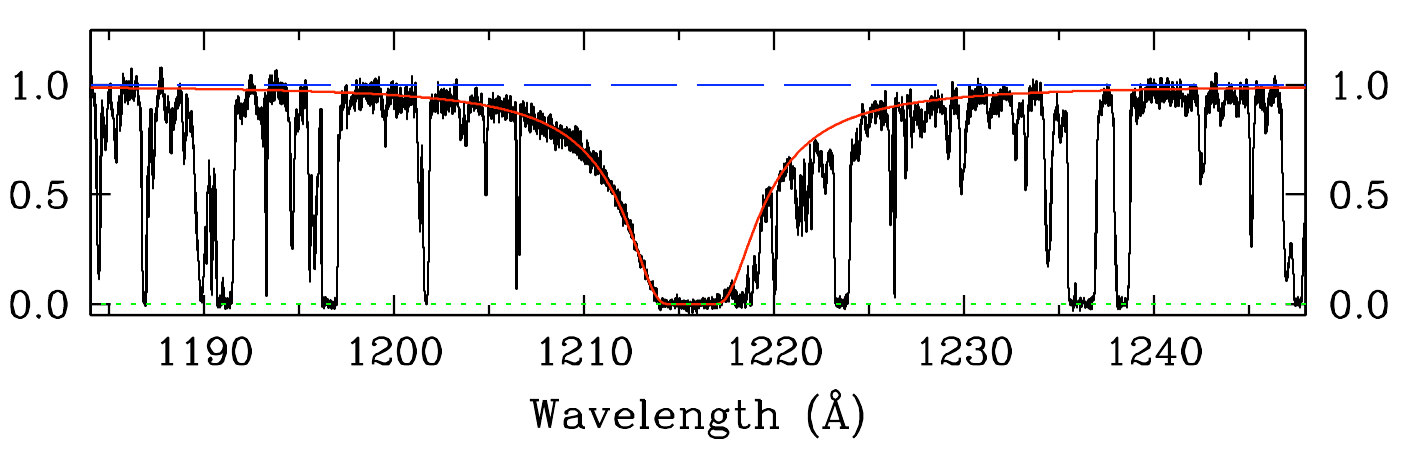}}
\vspace{-0.25cm}
\caption{Example of a damped Lyman alpha absorption system.
The strong Ly$\alpha$ absorption line at a rest wavelength
of 1216\,\AA\ (red line) is formed by high column densities of 
\textit{neutral} gas, which are normally found in or near galaxies.
Such strong spectral features are easily distinguished from the 
multitude of weaker absorptions due to more diffuse and ionised
gas in the intergalactic medium. The $y$-axis is flux relative to
the continuum of the background QSO against which the absorption
is seen.
}
\label{fig:DLA}
\end{figure}

QSOs are intrinsically
very bright and can thus be seen to very large distances; 
their spectra  provide a
backdrop against which the absorption lines produced 
by intervening gas, in galaxies and
the intergalactic medium, can be studied in much 
more detail than would otherwise be possible.
Among the different types of gas clouds 
that can produce absorption lines in QSO
spectra, we have learnt to recognise those 
most likely to be associated with galaxies still
at an early stage of evolution---the so-called 
`damped Lyman alpha systems',  or DLAs for short. 
These systems are easily identified, even in low resolution
spectra, from the strong Ly$\alpha$ absorption line at 1216\,\AA\ 
indicative of a high column density of neutral hydrogen,
$N$(H\,{\sc i})\,$\ge 2 \times 10^{20}$\,cm$^{-2}$.
An example is reproduced in Figure~\ref{fig:DLA}.
Nearly two  thousand are
now known thanks to 
large spectroscopic surveys of the sky which have been completed
in the last few years, such as the Sloan Digital Sky Survey.

High resolution spectroscopy of DLAs over the last 20 years
(e.g. [1] and [2] and references therein)
has characterised their metallicity distribution.
Most DLAs are metal-poor (unlike the brighter and more vigorously
star-forming Lyman break galaxies), with typical (i.e. median)
[Fe/H]\,$\simeq -1.5$. However, there is a tail in the distribution
extending to [Fe/H]\,$< -3$; it is these rare, least enriched DLAs
that best complement and extend studies of 
extremely metal-poor stars in our Galaxy.

The physics of the interstellar medium is much simpler than that
of stellar atmospheres. Consequently, deducing element abundances
from high resolution spectra of DLAs has several advantages over
its stellar equivalent. In particular, concerns about the physical
conditions under which the absorption lines are formed 
(i.e. LTE or NLTE) do not apply to the DLAs---all the absorption
lines used in abundance analyses are resonance lines from the
ground states of the relevant atoms and ions. Similarly, we are not
concerned with 1D vs. 3D modelling. 
Furthermore, when the gas is mostly neutral,
as is the case in DLAs, most elements are concentrated in
one dominant ionisation stage (usually either neutrals or
first ions) whose ionisation potential is greater than that 
of hydrogen. Thus, no allowances need to be made 
for unseen ion stages: `what you see is what you get'.
Finally, there is the obvious attraction of carrying out these 
abundance measurements at high redshift, when the Universe was
only 2--3\,Gyr old, and in a wider cosmological context than
the parochial perspective of our Local Group.

We have recently concluded the first 
survey for very metal-poor DLAs [3]. Our sample 
is only moderately large, consisting of 22 DLAs
with [Fe/H]\,$< -2$. On the other hand, with 
approximately one clear night on a 8--10\,m telescope
needed to secure spectra of high resolution
(FWHM\,$\simeq 7$\,km~s$^{-1}$) and moderately
high signal-to-noise ratio (S/N\,$\simeq 15$--50) per DLA,
the survey represents a considerable observational
effort over a number of years. In this conference report,
we highlight the most significant results uncovered
by the survey. In all cases we adopt the solar abundance scale
of Asplund et al. 2009 [4].

\section{Oxygen at low metallicities}

It has been known for nearly fifty years that
``Moderately metal-deficient stars... have a
large O/metals ratio.'' quoting the words
of Peter Conti and collaborators in 1967 [5].
Since then, the increasingly well-documented rise
in [O/Fe] as the iron abundances
decreases from solar values to 
[Fe/H]\,$ = -1$ has been a cornerstone
of Galactic chemical evolution models
(e.g. [6]). As larger telescopes brought
within the reach of high resolution spectroscopy
stars of lower and lower metallicity, a debate has ensued
as to whether the increase in [O/Fe] with decreasing
[Fe/H] continues monotonically to reach values
as high as [O/Fe]\,$\simeq +1$, or settles
on a roughly constant plateau at a more modest
[O/Fe]\,$\simeq +0.4$. The controversy
is due in part to differing values of the
oxygen abundance deducded from the analysis
of different spectral features, each suffering 
to different degrees from (at times uncertain)
corrections for NLTE and 3D effects.

\begin{figure}[]
\centerline{\includegraphics[width=0.5525\textwidth]{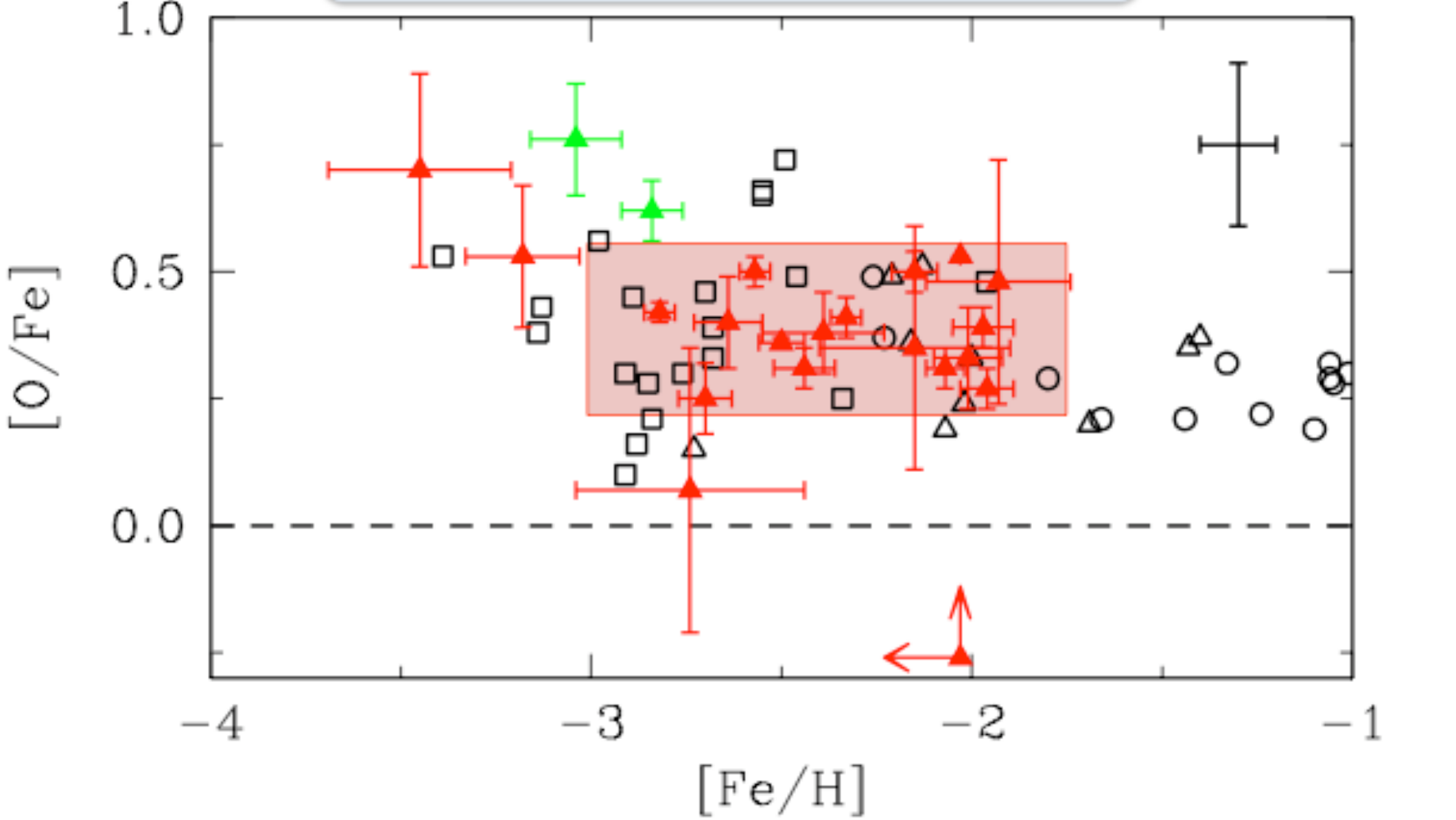}
\includegraphics[width=0.425\textwidth]{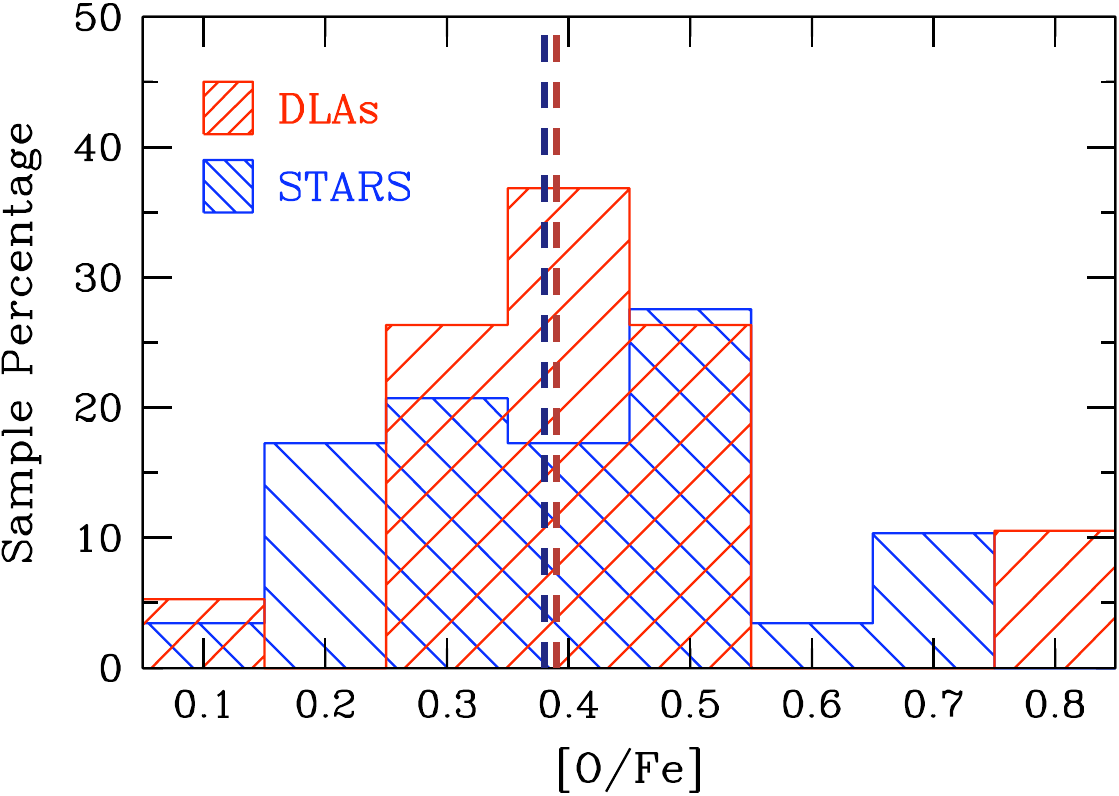}}
%\vspace{-0.25cm}
\caption{\textit{Left:} Oxygen enhancement at low metallicities in DLAs (filled 
symbols) and Galactic halo stars measured from the
forbidden [O\,{\sc i}]\,$\lambda 6300$ line (open symbols) . 
Original sources for the
stellar measurements can be found in [3].
\textit{Right:} The distributions [O/Fe] ratios in
DLAs and Galactic stars 
with ${\rm [Fe/H]} \leq -2$
are very similar when values in the
latter are deduced from the weak, but more reliable, 
[O\,{\sc i}]\,$\lambda 6300$ line. 
The oxygen enhancement is relatively modest, 
[$\langle {\rm O/Fe} \rangle$]\,$ \simeq +0.4$
in both high redshift gas clouds and in local, old stars.
}
\label{fig:O/Fe}
\end{figure}

These issues have now been largely settled by 
measurements of [O/Fe] in very metal-poor DLAs.
As can be seen from Figure~\ref{fig:O/Fe},
in DLAs with [Fe/H]\,$= -2$ to $-3$,
the [O/Fe] ratio is relatively uniform around
a moderate enhancement:
[$\langle {\rm O/Fe} \rangle$]\,$= +0.35 \pm  0.09$.
This mean value is in good concordance with those
found in old stars of the Milky Way within the same
metallicity interval, \textit{provided} the oxygen abundance
is deduced from the weak [O\,{\textsc{i}]\,$\lambda 6300$ line.
Among the various O spectral features targeted in the metal-poor
regime, this is the line which forms in LTE and is subject to only
modest 3D corrections.

Does the [O/Fe] ratio remain constant at the above value
when the metallicity is lower than 1/1000 of solar, or
does it climb to more pronounced enhancements?
This is an important question, closely linked
to the initial mass function (IMF)
of the first few generations of stars. Unfortunately,
we do not have an answer yet. There are 
very tentative indications of a further increase
(see Figure~\ref{fig:O/Fe}), but the number of known
DLAs with such extremely low metallicities
is still very small, while the stellar
[O\,{\textsc{i}]\,$\lambda 6300$ line
becomes too weak to be measured reliably
with current instrumentation. 
Both obstacles will undoubtedly
be overcome in the not too distant future,
so watch this space!

\section{Carbon at low metallicities}

The relative abundances of carbon and oxygen
exhibit a complex behaviour as a function of
metallicity (in this case measured via O/H).
Referring to Figure~\ref{fig:C/O}, it can be
seen that in Galactic stars C becomes progressively less
abundant than O as the O abundance
drops from solar to 1/10 solar;
at [O/H]\,$\simeq -1$,  [C/O]\,$\simeq -0.5$.
This behaviour, which has been known for a while,
is thought to arise from the combination of two effects:
the reduced mass loss from massive stars at 
lower metallicities and the delayed production
of C from the late stages in the evolution of intermediate-
and low-mass stars (e.g. [7]).

\begin{figure}[]
\centerline{\includegraphics[width=0.8\textwidth]{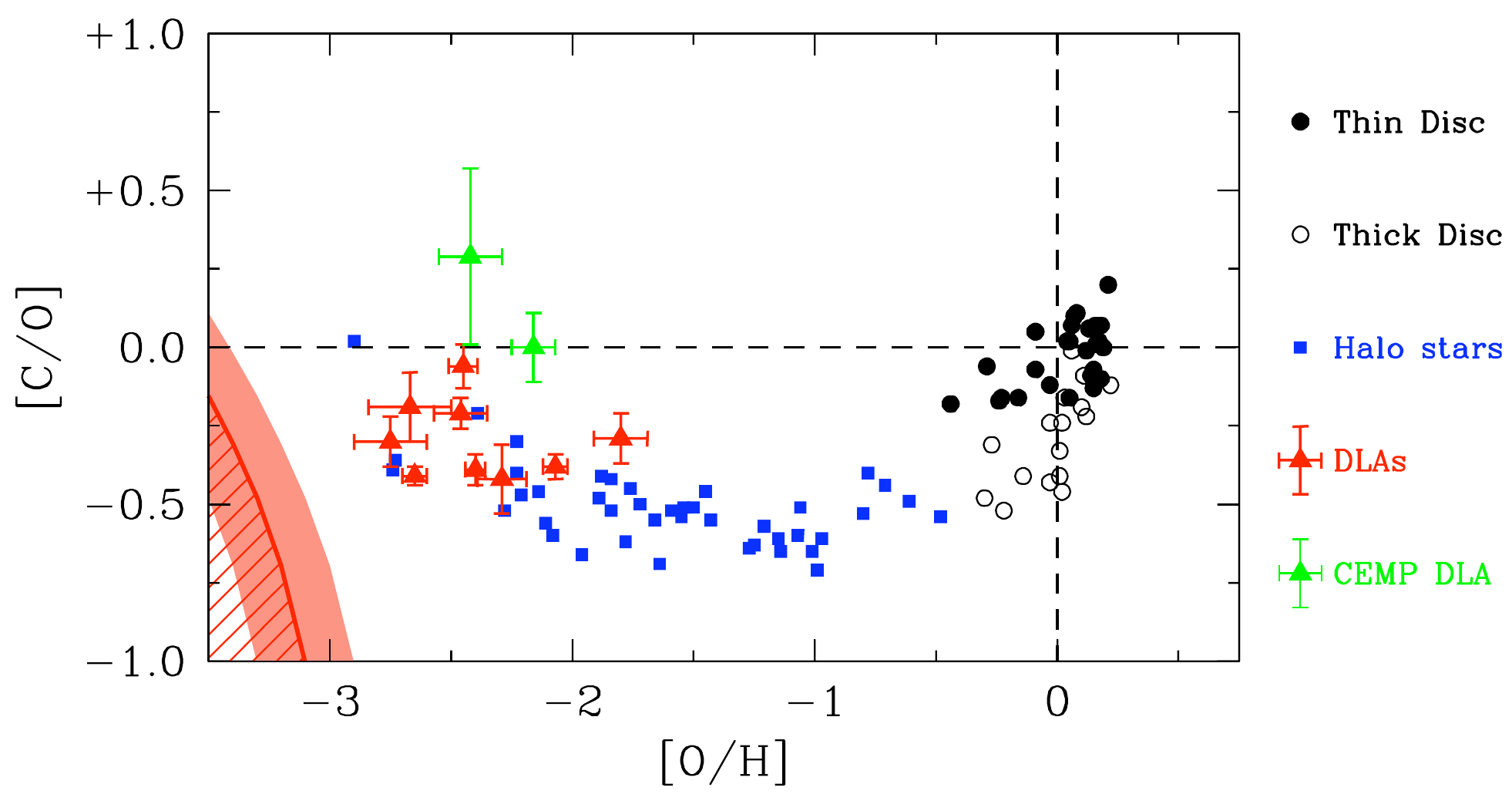}}
\vspace{-0.15cm}
\caption{Carbon and oxygen abundances
 in DLAs and Galactic stars as indicated. 
Original sources for the
stellar measurements can be found in [3].
}
\label{fig:C/O}
\end{figure}

What had not been appreciated until relatively recently
is that, as we move to even lower metallicities,
the C/O ratio appears to slowly recover, reaching again
near-solar proportions when [O/H]\,$\sim -3$
(see Figure~\ref{fig:C/O}).
This behaviour, first seen in metal-poor halo stars by Akerman et al. [7],
was totally unexpected---without an additional source
of C at the lowest metallicities, the C/O ratio was 
predicted to plummet when [O/H]\,$\aplt -2$.
For this reason, Akerman et al. `hedged their bets',
by flagging the possibility that the trend may not
be real, but be caused by the increasing importance
with decreasing metallicity of NLTE corrections to the  
C\,{\sc i} and O\,{\sc i}  lines used in their abundance determinations.
Such concerns have now  been  largely assuaged
by our finding a good agreement between the [C/O]
values measured in DLAs and Galactic stars
with [O/H]\,$< -2$. A concerted analysis of the 
NLTE corrections [8] has also confirmed that the stellar 
rise is indeed real.

It thus seems to be reasonably well established that
at the lowest metallicities stellar nucleosynthesis
can produce copious amounts of C, a conclusion which 
we suspect must be in some way linked to increasing
proportion of C-enhanced metal-poor (CEMP) stars at the 
lowest metallicities. This topic has been
discussed extensively at this meeting (see contributions by
Carlo Abate, Wako Aoki, and Catherine Kennedy).
Of particular relevance to our understanding of the nature
of CEMP stars is the discovery of one, or possibly two,
CEMP DLAs in the course of our survey [9], [10].
In the DLAs,  the C-enhancement presumably  reflects
the overall chemical composition of the gas from 
which subsequent generations of stars formed,
rather than being the result of mass-transfer from
an unseen companion. Thus, these rare CEMP DLAs
may well be the high-redshift counterparts of the 
CEMP-no stars found in the halo of the Galaxy 
and its companions.  Again, the connection will
undoubtedly be clarified as more metal-poor DLAs are
identified and targeted with high resolution spectroscopy.

\section{Nitrogen at low metallicities}

With two UV triplets available and an ionisation potential
close to that of H, N is an element whose abundance
is relatively easy to measure in DLAs. 
This is  fortunate given
the difficulty of analogous measurements in metal-poor stars
and the lack of H\,{\sc ii} regions (the other source of N abundance
determinations)  with metallicities below 1/30 of solar.
Thus, it is thanks to DLAs that we have been able to probe the 
nucleosynthesis of N in the very metal-poor regime 
(Figure~\ref{fig:N/O}).

\begin{figure}[h!]
\medskip
\centerline{\includegraphics[width=0.75\textwidth]{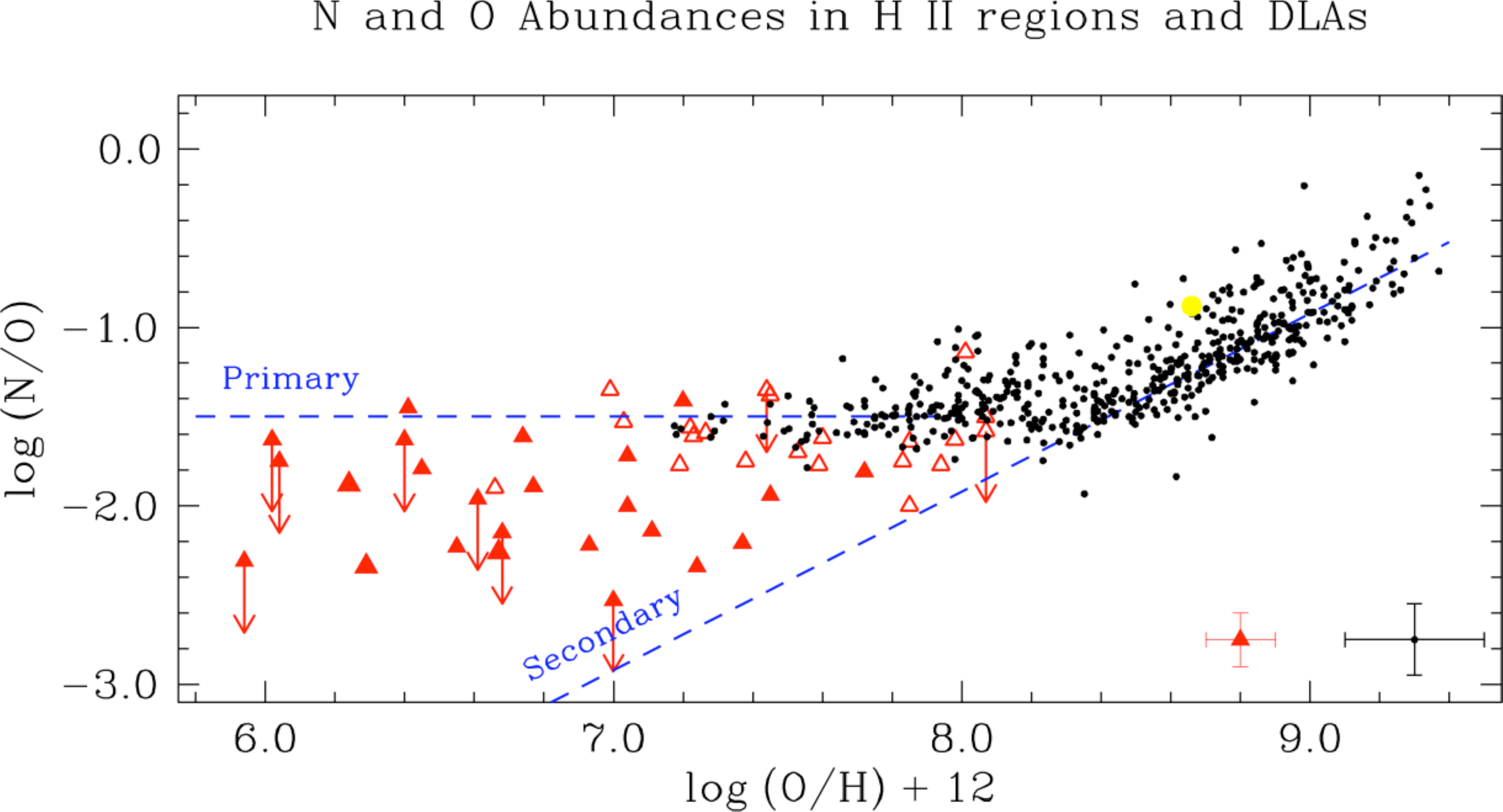}}
%\vspace{-0.15cm}
\caption{Nitrogen and oxygen abundances
 in DLAs (triangles) and nearby H\,{\sc ii} regions (small dots).
Filled triangles denote DLAs in which O/H could be measured
directly, while open triangles are cases where the available
O\,{\sc i} lines are saturated and the oxygen abundance
was deduced from S/H, assuming a solar S/O ratio. 
Original sources for the
H\,{\sc ii} region measurements can be found in [10].
The error bars in the bottom right-hand corner give an indication of the typical
uncertainties. The large yellow dot shows the solar abundances
of the two elements.
}
\label{fig:N/O}
\end{figure}

In a plot of log (N/O) vs. log (O/H), metal-poor DLAs 
fall between the boundaries set by the primary 
and secondary production of N.
The two terms denote the synthesis of N from 
seed C and O which are either manufactured 
by the star during He burning (primary N), or 
were already present when the star 
first condensed out of the interstellar medium (ISM; secondary N).
This is qualitatively in agreement with the idea
(e.g. [12])
that the main sources of primary N are 
intermediate-mass stars on the asymptotic giant branch,
such that the release of nitrogen into the ISM  takes place some time 
after the massive stars which are the main producers 
of O have exploded as Type II supernovae.

A few additional points are noteworthy.
First, the finding that the DLA measures are spread
approximately uniformly between the 
primary and secondary boundaries suggests that
the release of N into the ISM takes place over longer timescales
than the few $10^8$ years normally considered
between the release of N and O, unless we
observe most DLAs at a special time, which seems unlikely
(recall that the age of the Universe at redshifts $z = 2$--3,
where most DLAs are observed, is $\sim 2$--3\,Gyr).
In any case, the survey results dispel the idea
of a bi-model distribution in the values of N/O
which had been mooted on the basis of smaller samples
of metal-poor DLAs [13].
Second, there seems to be a `floor' to the minimum value
of the N/O ratio at $\log {\rm (N/O)} \simeq -2.3$.
In most cases, this is close to the limit of the DLA
measurements, so that it is unclear whether the 
`floor' simply reflects the sensitivity of our current
instruments or is a real minimum value of the relative
abundances of these two elements. If the latter were true,
the simplest interpretation would be that this is 
primary N manufactured by \textit{massive} stars,
a much debated possibility.

\section{Primordial Deuterium}

\begin{figure}[b!]
\centerline{\includegraphics[width=0.85\textwidth]{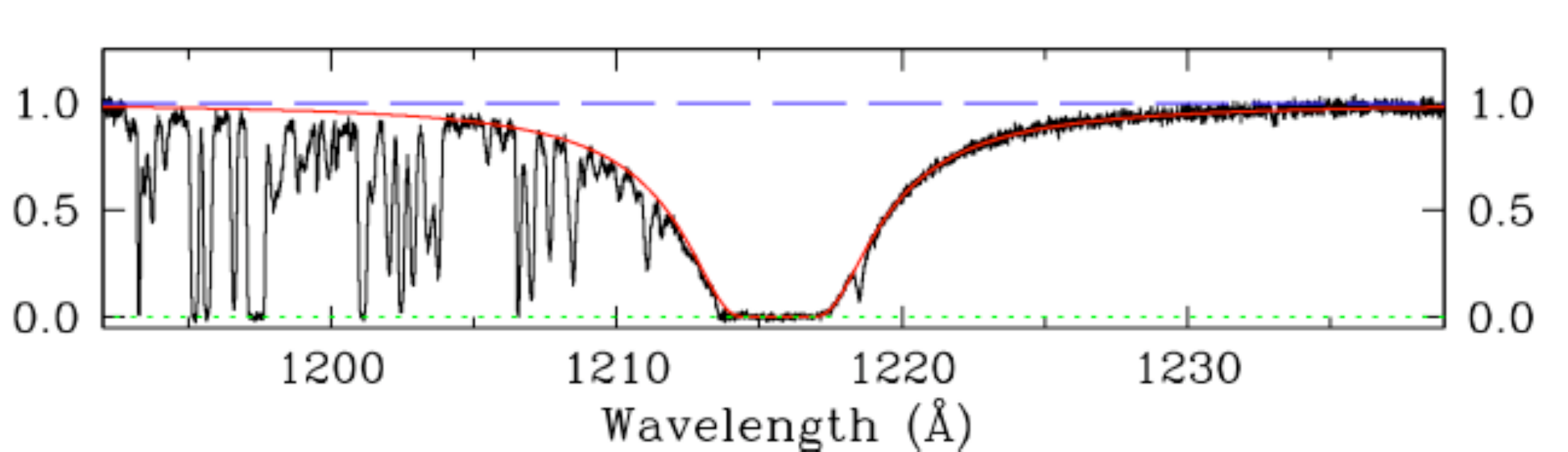}}
\centerline{\includegraphics[width=0.85\textwidth]{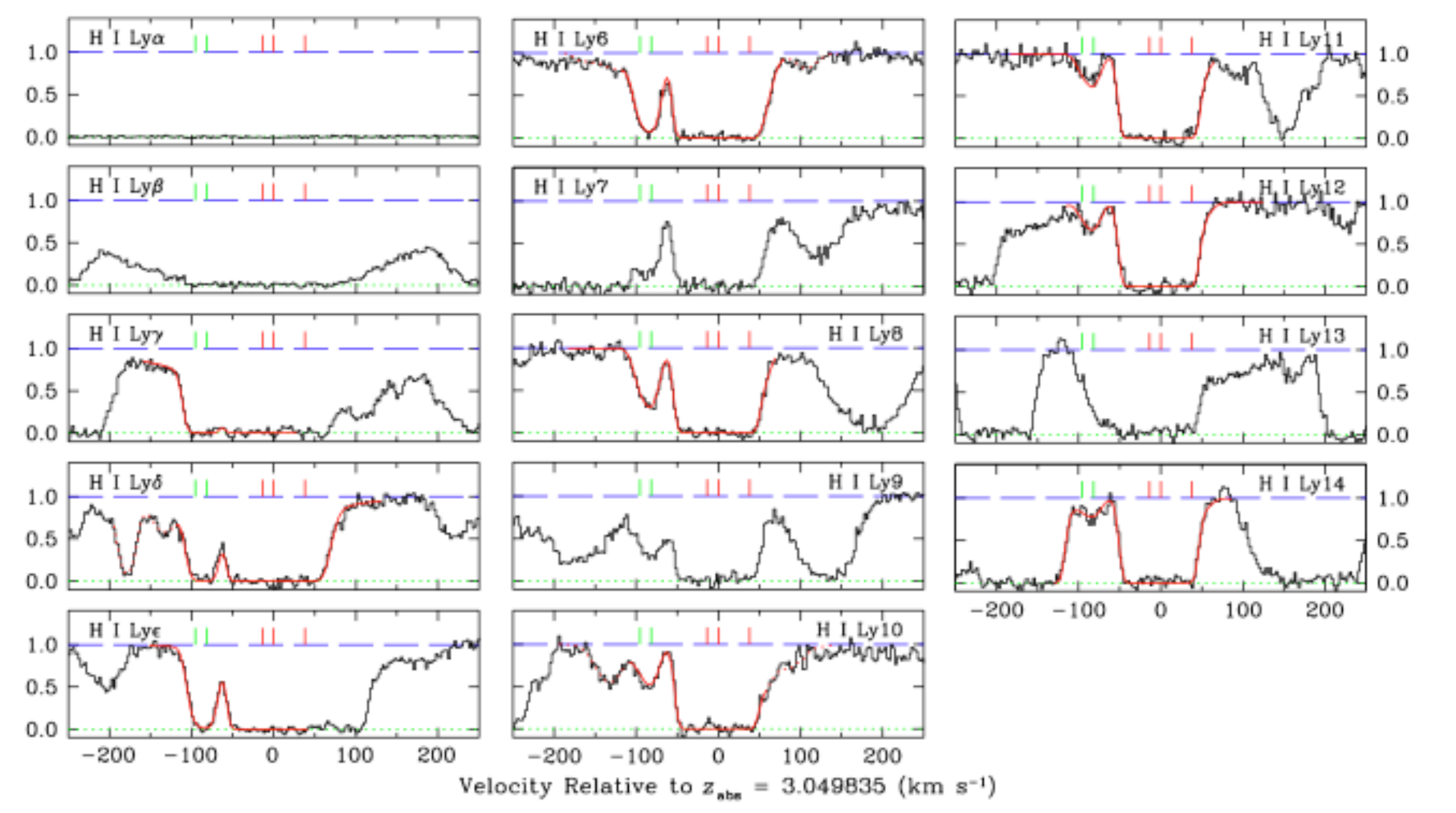}}
%\vspace{-0.15cm}
\caption{\textit {Top panel:} The Ly$\alpha$ region in the UVES
spectrum of SDSSJ1419$+$0829. The spectrum (black histogram)
has been normalised to the QSO continuum and reduced to the rest-frame of 
the $z_{\rm abs} = 3.04984$ DLA. The red line
shows the best-fitting damped Ly$\alpha$ absorption
profile, produced by a neutral hydrogen column density
$\log N{\rm (H\,\textsc{i})/cm}^{-2} = 20.391 \pm 0.008$.
\textit{Bottom panels:} Lyman series lines in the DLA.
The black histogram is the observed spectrum, while the red
continuous line is the model fit to the absorption features.
Vertical tick marks above the spectrum indicate the three 
absorption components contributing to the H\,{\sc i} (red) and
D\,{\sc i} (green) absorption. In all panels, the $y$-axis scale is residual intensity.}
\label{fig:1419}
\end{figure}

The most metal-poor DLAs are a window not only on early stellar
nucleosynthesis, but also on \textit{primordial} nucleosynthesis
in the first few minutes after the Big-Bang (BBN). In particular,
such DLAs are the best astrophysical environments for the determination
of the primordial abundance of deuterium, (D/H)$_{\rm p}$, 
`the baryometer of choice' in the words of  David Schramm
and Mike Turner [14]. Consider the following.
For metallicities lower than 1/100 of solar, the corrections
for the astration of deuterium (its destruction through cycles
of star formation) are minimal. The most metal-poor DLAs are
also the ones with the simplest internal kinematics; the 
absorption usually takes place in just  one or two
discrete components with internal velocity dispersion
of only a few km~s$^{-1}$, allowing the D 
isotope shift of $-82$\,km~s$^{-1}$ to be resolved.
Furthermore, the high column densities of neutral gas in DLAs
make \textit{many} absorption lines in the Lyman series
of D\,{\sc i} accessible to observation and measurement,
greatly improving the accuracy with which the D/H ratio
can be determined, compared to other classes of QSO absorbers.

Reliable measures of (D/H)$_{\rm p}$ are still scarce after nearly
twenty years of high resolution spectroscopy with 8--10\,m 
class telescopes, but our survey of VMP DLAs has the potential
of significantly improving current statistics. One of our targets, 
the $z_{\rm abs} = 3.04984$ DLA in the QSO J1419+0829,
turns out to have near-ideal properties 
for an accurate determination of (D/H)$_{\rm p}$.
This realisation spurred us to develop software specifically 
designed to deduce the best fitting value of D/H and to 
assess comprehensively the random and systematic errors 
affecting this determination [15].  In Figure~\ref{fig:1419} we
have reproduced portions of the VLT-UVES spectrum of this
QSO encompassing lines in the Lyman series of the DLA.
D\,{\sc i} absorption is clearly separated from nearby H\,{\sc i}
in eight transitions of widely different $f$-values, from 
Ly$\delta$ to Ly14.

\begin{figure}[t!]
\centerline{\includegraphics[width=0.55\textwidth]{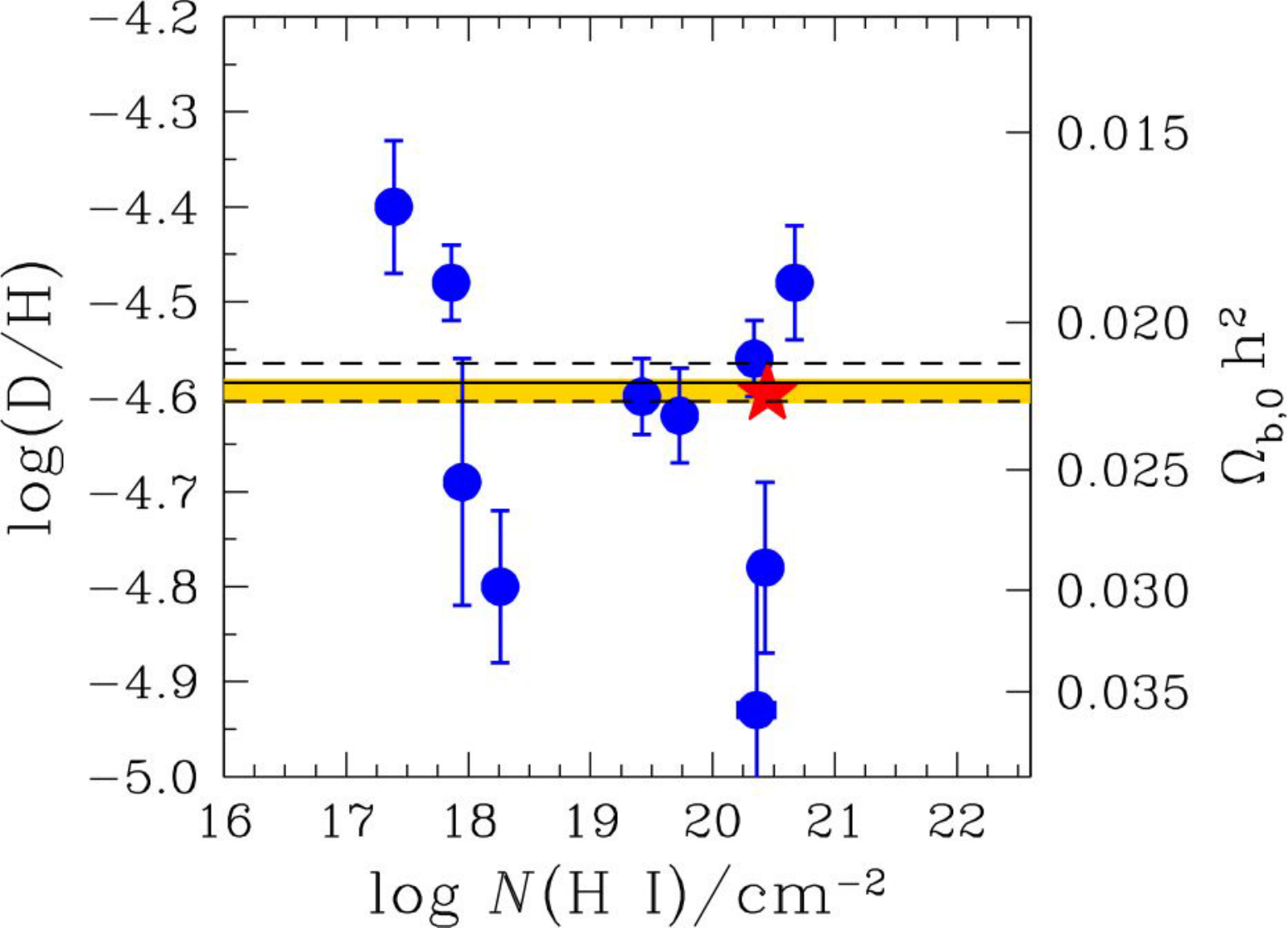}}
%\vspace{-0.15cm}
\caption{Measures of the deuterium abundance in high redshift 
QSO absorbers. Only cases were the deuterium absorption is 
clearly resolved from nearby spectral features are shown. 
The red star refers to the most recent measurement by Pettini \& Cooke [15], 
with errors smaller than the symbol size. 
The sources of earlier determinations are given in [15].
The horizontal lines 
are drawn at the weighted mean value of log (D/H) and its error, 
as determined with the bootstrap method. 
The yellow shaded area shows the range in	
$\Omega_{\rm b, 0} h^2 {\rm (CMB)}$ from [16].
}
\label{fig:D}
\end{figure}

We find ${\rm D/H} = (2.535 \pm 0.05) \times 10^{-5}$, 
where the uncertainty includes both random and systematic
errors estimated via Monte Carlo techniques.
This is the most precise estimate
of (D/H)$_{\rm p}$ to date (see Figure~\ref{fig:D}).
It implies 
$\Omega_{\rm b, 0} h^2  = 0.0223 \pm 0.0009$\footnote{
Note that the error has doubled from 2\% to 4\% in 
converting (D/H)$_{\rm p}$ to $\Omega_{\rm b, 0} h^2$.
The additional source of error is the difference 
between  theoretical and experimental estimates of  the
nuclear reaction rate $d(p,\gamma)^3{\rm He}$ [17].},
in very good agreement with 
$\Omega_{\rm b, 0} h^2{\rm (CMB)}  = 0.0222 \pm 0.0004$
deduced from the analysis of the power spectrum
of the temperature fluctuations of the cosmic microwave
background [16].

The consistency between the cosmic density of baryons measured
from entirely different physical processes operating,
respectively, in the first few minutes of our Universe's existence
and $\sim 375\,000$ years later gives confidence in the validity of
the standard cosmological model. As has been discussed extensively
in recent years (e.g. [18]), the comparison between
$\Omega_{\rm b, 0} {\rm (BBN)}$ and $\Omega_{\rm b, 0} {\rm (CMB)}$
can be used to put constraints on the number of sterile neutrinos
(sometimes fashionably referred to as `dark radiation')
and on a possible lepton asymmetry.
While these intriguing speculations have yet to be put to rest,
we note here that what we consider to be the most accurate
estimate of the primordial abundance of D limits the
number of neutrino families to ${\rm N}_\nu = 3.0 \pm 0.5$.

\section{Conclusions}

The role that DLAs can play in exploring the origin of
chemical elements has become increasingly appreciated
since the first efforts in this direction (e.g. [19]).
The survey of the most metal-poor DLAs we have described 
here has resolved a number of issues lingering from
analogous stellar studies. 
The enhancement of O relative to Fe
is relatively modest, amounting to a factor of $\sim 2.5$
in the metallicity range [Fe/H]\,$= -2$ to $-3$. 
It seems likely that the `First Stars' were copious C producers.
Our ideas about the relative importance of primary
and secondary production of N appear to be approximately 
correct. The abundance of D measured in high redshift
clouds confirms the fundamental framework of 
Big-Bang nucleosynthesis 
without recourse to non-standard physics.

As is always the case, these results whet our appetite
for further questions. Does the O-enhancement increase
at the lowest metallicities, as may be expected if the 
First Stars were very massive? Can we find further
examples of CEMP DLAs, and can we constrain the IMF
of the First Stars from their chemical composition?
Can massive stars produce primary N, perhaps through
rotation? Is it possible to improve the overall
accuracy of  $\Omega_{\rm b, 0} {\rm (BBN)}$
and thereby place tighter constraints on non-standard
physics in conjunction with the forthcoming
measurements of $\Omega_{\rm b, 0} {\rm (CMB)}$
from the Planck mission? We are confident that significant
progress will have been made on at least some of these
issues by the time we meet again at the next
\textit{`Nuclei in the Cosmos'} meeting.\\

It is a pleasure to acknowledge our collaborators in the projects
described in this conference presentation: Regina Jorgenson,
Michael Murphy, Poul-Erik Nissen, Gwen Rudie, and
Chuck Steidel. We are grateful to the 
time allocation panels of the Keck and VLT telescopes
for their support of this demanding observational programme.
Special thanks the organising committee of \textit{Nuclei in the Cosmos XII}
for running a smooth, engaging, and altogether very successful
meeting.

\vspace{0.75cm}

\noindent {\bf Questions}

\bigskip

\noindent  \textit{Jennifer Johnson}:  What are the sizes and nature of the systems 
that you are probing and can that tell us anything about mixing in the early Galaxy?

\smallskip

\noindent  \textit{Max Pettini}:  It is a source of considerable frustration that, while
we can study in great detail the physical properties of these absorbers from high
resolution spectroscopy, it has proved very difficult to associate them with
particular classes of galaxies. Imaging searches for their optical counterparts
have so far shown that a wide range of galaxies, spanning many magnitudes
in the galaxy luminosity function, can give rise to DLAs. My instinct tells me
that the most metal-poor DLAs, on which I have concentrated today, may also
be the most difficult to image directly, if they have indeed experienced very low
levels of star formation as their chemically unevolved status suggests. \\

\noindent  \textit{Chiaki Kobayashi}: Can you comment on other problems in 
observations eg. dust depletion and temperature effect?

\smallskip

\noindent  \textit{Max Pettini}:  Dust depletions are very much reduced with
decreasing metallicities and are expected to make only negligible corrections 
to the element abundances deduced from the gas-phase in VMP DLAs.
We can get a good handle on the temperature of the gas by comparing the
widths of absorption lines from elements of different masses. Temperatures 
of a few thousand degrees are indicated for the DLAs I have focussed on today. \\

\noindent  \textit{Taka Kajino}:  The analysis of WMAP7 data of CMB anisotropies 
alone gives us ${\rm Y}_{\rm p} \geq  0.3$ and ${\rm N}_\nu \geq 5$, 
which are too large to accept. Is it possible to detect $^4{\rm He}$ abundance 
from Lyman-alpha clouds as you did for D?

\smallskip

\noindent  \textit{Max Pettini}:  I wish it were possible! But the He\,{\sc i} lines
all occur at very short wavelengths, where the Universe is opaque to EUV radiation.
In the words of the sage: ``Never say never again'', but the difficulties are enormous. \\

\noindent \textit{Harriet Dinerstein}: You mentioned that the C-enhanced DLAs 
may be the counterparts of the CEMP-no stars. 
Do you have any upper limits on neutron-capture elements in these systems?

\smallskip

\noindent  \textit{Max Pettini}: This is a very good point, Harriet. To really substantiate that 
statement we would need to measure the abundances of neutron-capture elements,
a task for the 30-m class telescopes of the future. \\

\noindent  \textit{Gabriele Cescutti}: What are the results concerning  nitrogen in DLA systems?

\smallskip

\noindent  \textit{Max Pettini}: I didn't have sufficient time to discuss nitrogen in my
oral presentation, but I have included the results you are interested in in Section 4 of this
conference report.

\end{document}